\documentclass[twocolumn,showpacs,preprintnumbers,aps,amsmath,amssymb,
floatfix,superscriptaddress]{revtex4-1}

\usepackage{epsfig}
\usepackage{graphicx}
\usepackage{dcolumn}
\usepackage[usenames]{color}

\begin{document}

\newcommand{\jpt}{J$^{\pi } $=2$ ^{+} $}
\newcommand{\alphaD}{\alpha_{{}_{\hspace{-0.5pt}D}}}

\title{Fine structure of the pygmy quadrupole resonance in $^{112,114}$Sn isotopes}

\author{N.~Tsoneva}
\altaffiliation{Institut f\"ur Theoretische Physik, Universit$\ddot{a}$t Gie$\beta$en,
Gie$\beta$en, D-35392, Germany}
\affiliation{Extreme Light Infrastructure (ELI-NP) \& IFIN-HH, Horia Hulubei National
Institute of Physics and Nuclear Engineering, Bucharest-M\u{a}gurele, Romania}

\author{M.~Spieker}
\altaffiliation{present address: NSCL, Michigan State University, 640 South Shaw Lane, East Lansing, MI
48824, USA}
\affiliation{Universit\"at zu K\"oln, Institut f\"ur Kernphysik, K\"oln, D-50937, Germany}

\author{H. Lenske}
\affiliation{Institut f$\ddot{u}$r Theoretische Physik, Universit$\ddot{a}$t Gie$\beta$en,
Gie$\beta$en, D-35392, Germany}

\author{A. Zilges}
\affiliation{Universit\"at zu K\"oln, Institut f\"ur Kernphysik, K\"oln, D-50937, Germany}

\date{\today}

\begin{abstract}
The electric quadrupole response in $^{112,114}$Sn isotopes is investigated by energy-density functional (EDF) and three-phonon quasiparticle-phonon model (QPM) theory with special emphasis on 2$^+$ excitations located above the first collective quadrupole state and below 5 MeV. Additional quadrupole strength clustering as a sequence of states similar to the recently observed pygmy quadrupole resonance in $^{124}$Sn is found. The spectral distributions and transition densities of these 2$^+$ states show special features being compatible with oscillations of a neutron skin against the isospin-symmetric nuclear core. Furthermore, two new ($p$, $p' \gamma$) Doppler-shift attenuation (DSA) coincidence experiments were performed at the SONIC@HORUS setup. Quadrupole states with excitation energies up to 4.2 MeV were populated in $^{112,114}$Sn. Lifetimes and branching ratios were measured allowing for the determination of the reduced quadrupole transition strengths to the ground state. A stringent comparison of the new data to EDF+QPM theory in  $^{112}$Sn and $^{114}$Sn isotopes hints at the occurrence of a low-energy quadrupole mode of unique character which could be interpreted as pygmy quadrupole resonance. 
\end{abstract}

\maketitle

Investigations of nuclear structure in hitherto inaccessible mass regions far from the valley of $\beta$-stability are becoming a reality nowadays at rapidly developing radioactive beam facilities\,\cite{Muenz01,Blum13,Bor16,Bal16,Gad16,Nav11,Kub92}. Similarly to the halos in light nuclei\,\cite{Tan1-85,Tan2-85} the prevalence of one type of nucleons in medium and heavy mass nuclei may result in the formation of a nuclear skin, containing either pure proton or neutron matter, respectively. The dynamics of such nuclear systems is very sensitive to the nuclear symmetry energy and its density dependence \,\cite{Ton17} which on the other hand is imprinted in a static mean-field observable - the neutron-skin thickness\,\cite{Ton17,Pie06,Pie11,Tam11}. In case, of moderate and heavy mass nuclei, the latter is defined  as the difference between the neutron and proton root-mean-square (rms) radii $\delta r=\sqrt{<r^2>_n}-\sqrt{<r^2>_p}$. Furthermore, the presence of a neutron skin may influence the nuclear response on external electromagnetic and hadronic fields. An observable which could be sensitive to induced skin effects on nuclear excitations, especially at low-energies, is the dipole polarizibility. The latter is proportional to the second moment of the dipole photoabsorption cross section and it could be also useful for constraining neutron skin and symmetry energy\,\cite{Ton17,Tam11}, nuclear matter, the equation of state and the properties of neutron stars\,\cite{Yos07,Hor01}. 

Recently, new modes of excitation closely connected with the skin phenomena - the pygmy dipole resonance (PDR)\,\cite{Sav13,Tso04,Vol06,Tso08,Sch08,Ton10,Sch13,Bra15} and first hints for its natural extension at higher multipolarities - the pygmy quadrupole resonance (PQR)\,\cite{Tso11,Pel15,Spi16} have been observed. During the last years the PDR was the subject of many experimental and theoretical investigations\,\cite{Sav13,Bra15,Paa07}. 

In many studies a close correlation between the total PDR strength and the neutron or proton skin thickness was obtained\,\cite{Pie11,Tso04,Vol06,Tso08,Sch08,Ton10,Sch13} and it was found that the PDR might affect directly (n,$\gamma$) reactions cross sections and rates of the s- and r- process synthesis of heavier nuclei in stellar environments\,\cite{Gor02,Tso15}. 
The existence of a PQR originally predicted theoretically in \,\cite{Tso11} was recently supported by independent experiments in $^{124}$Sn using different probes and techniques\,\cite{Pel15,Spi16}. In these studies, it is obtained that the PQR resembles the characteristics of the PDR and could be related to higher-order multipole oscillations of weakly-bound neutron or proton excess nucleus forming a neutron or a proton skin, respectively. 
The newly discovered nuclear PQR mode is not yet investigated in such detail as the meanwhile well established  PDR excitation. The few existing theoretical studies suggest a strong mixture of skin, shell, pairing and core polarization contributions on low-energy quadrupole excitations and in particular in the PQR\,\cite{Tso11,Pel15,Spi16}. Furthermore, with the increase of the collectivity of the quadrupole states the reliable distinction of the pure neutron or proton quadrupole skin oscillations seems to be a difficult task, especially for nuclei with small or moderate neutron excess.
Thus, studies of low-energy 2$^+$ states and their theoretical interpretations could be even more sensitive to the choice of models and interactions than for the PDR mode. Applications of the theory to different types of spectroscopic data are the appropriate way to identify clearly the PQR mode and to verify the predicted properties unambiguously.

The aim of this paper is to examine the dependence of the PQR mode on the charge asymmetry. 
As a suitable case we investigate low-energy quadrupole excited states in tin isotopes $^{112}$Sn (with 12 valence neutrons and N/Z=1.24) and $^{114}$Sn (with 14 valence neutrons and N/Z=1.28), respectively. Despite the relatively small neutron excess, it is expected that these nuclei develop a neutron skin \,\cite{Tso08}, and as a consequence additional low-energy quadrupole strength related to neutron skin oscillations \,\cite{Tso11}. For that purpose, we performed new elaborated theoretical and experimental studies in these nuclei. In addition, for the first time we investigated in detail the mechanism of the fragmentation of the PQR strength and the role of the multi-phonon coupling. This includes a refined theoretical analysis of the multi-phonon structure of the low-energy quadrupole excitations. Furthermore, the theoretical results and experimental data on branching ratios give a new insight into the understanding of the fine structure of the PQR and nuclear excitations.

\begin{figure}[tb]
\centering
\includegraphics*[width=84mm]{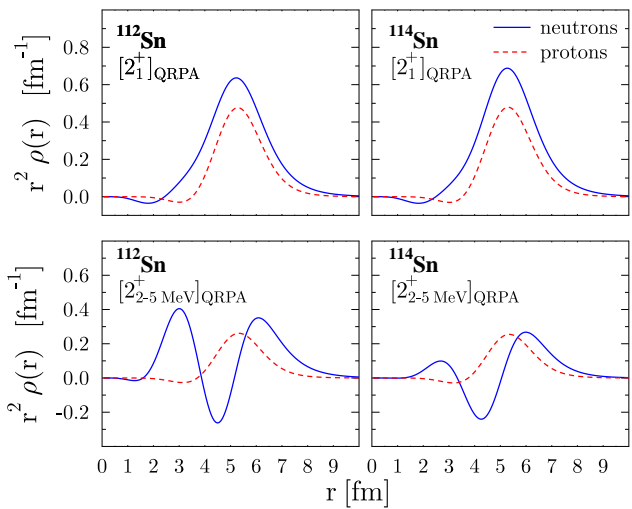}  
\caption{(color online) (top) Neutron and proton transition densities of the [2$^+_1$]$_{QRPA}$ states in $^{112}$Sn and $^{114}$Sn. (bottom) Summed neutron and proton transition density for the [2$^+$]$_{QRPA}$ states between 2 MeV and 5 MeV in $^{112}$Sn and $^{114}$Sn, respectively.}
\label{FIG1}
\end{figure}

The theoretical investigations are based on an advanced microscopic approach incorporating self-consistent energy-density functional (EDF) theory and the quasiparticle-phonon model (QPM)\,\cite{Sol76} including up to three-phonon couplings. A detailed description of our theoretical method is presented in our review paper\,\cite{Tso16} and its applications to a large variety of nuclear research are reported in \,\cite{Tso04,Vol06,Tso08,Sch08,Ton10,Sch13,Tso11,Tso15,Rus13,Kri15,Banu14,Cos15}.  

\begin{table}[htbp]
 \caption[]{B(E2) transitions of  $2^{+}_{1}$, $2^{+}_{2}$, $2^{+}_{3}$ and $4^{+}_{1}$ states in $^{112}\mathrm{Sn}$ and $^{114}$Sn.}
\centering
\begin{tabular}{llrrrrrrr}
\hline \hline \\
$J^{\pi}_{i}$&    & E $[$MeV$]$ &   &$I^{\pi}_{j}$ & $B(E2;$ & $J^{\pi}_{i}\longrightarrow I^{\pi}_{j})$ & [$\mathrm{W.u.}$]& \\
\hline \\
 & Exp.& EDF+QPM &   &  & Exp. & EDF+QPM &  \\
\hline \\
  &      & $^{112}$Sn   &             &          &     \\
\hline \\ 
$2^{+}_{1}$ & 1.26 & 1.20   &   & $0^{+}_{1}$ & 12.5(8)  & 12.1\\
$2^{+}_{2}$ & 2.15 & 2.45   &   & $0^{+}_{1}$ & - & 0.1\\
            &           &   &   & $2^{+}_{1}$ & 18(5) & 25 \\ 
$2^{+}_{3}$ & 2.48 & 2.79   &   & $0^{+}_{1}$ & 0.082(13) & 0.015\\
            &           &   &   & $2^{+}_{1}$ & 0.23(6)   & 0.92  \\ 						
            &           &   &   & $2^{+}_{2}$ &  -       & 0.06\\
$4^{+}_{1}$ & 2.25 & 2.19   &   & $2^{+}_{1}$ & 5.6(8)   & 11.4\\
\hline \\ 
  &      & $^{114}$Sn \\
\hline \\ 
& Exp.& EDF+QPM & IBM-2 &  & Exp. & EDF+QPM & IBM-2 \\
\hline \\ 
$2^{+}_{1}$ & 1.30 & 1.31   & 1.30   & $0^{+}_{1}$ & 11.1(7)  & 12.0 & 11 \\
$2^{+}_{2}$ & 2.24 & 2.56   & 2.46   & $0^{+}_{1}$ & $\leq$ 0.12 & 0.002 & 0.04\\
            &           &   &        & $2^{+}_{1}$ & $\leq$ 8 & 13 & 2 \\ 
$2^{+}_{3}$ & 2.45 & 2.85   & 2.54   & $0^{+}_{1}$ & 0.023(9) & 0.012 & 0.004\\
            &           &   &        & $2^{+}_{1}$ & 3(2)     & 3  & 17\\ 						
            &           &   &        & $2^{+}_{2}$ &  -       & 0.04 & 8 \\
$4^{+}_{1}$ & 2.19 & 2.15   & 2.28   & $2^{+}_{1}$ & 5.9(5)   & 7.9 & 19 \\
\hline \hline \\
\end{tabular}
\label{114Sn}
\label{Tab1}
\end{table}

The calculations are based on previously derived parameter sets \,\cite{Tso08,Tso11} but accounting for minor changes of the interaction parameters because of refined experimental results indicating a decreased collectivity of the first 2$^+$ states in the Sn isotopes. \,\cite{Jun11}. 

The dominance of neutron or proton transitions in nuclei implies a strong mixture of isoscalar and isovector contributions because it is related to the particle projector, $P_q$ = $\frac{1}{2}(1 \pm\tau_3)$, where $\tau_3$ = $\pm1 $ for neutrons and protons, respectively. The latter implies a pure isospin character only for N = Z nuclei \cite{Tso11,Pel15}. In the other cases one has an admixture of an isoscalar with a nonzero isovector transition amplitude. In general, the dominating term is the isoscalar one and it decreases with $(N-Z)/A$\,\cite{Tso11}.
The isospin effect on the QRPA phonon states could be examined via isoscalar and isovector multipole matrix elements which in the Sn isotopes are discussed in \,\cite{Tso11,Pel15,Spi16}. The theoretical observations show that most of the low-energy quadrupole strength has an isoscalar origin but also there are non-negligible contribution of isovector quadrupole strength. 

Furthermore, the presence of isoscalar and isovector contributions to the PQR excitations in $^{112,114}$Sn nuclei could be seen as well from the spacial distribution of neutron and proton QRPA transition densities which are shown in Fig.\,\ref{FIG1}. At excitation energies E$_x$ = $2-5$ MeV the main part of the oscillations is coming from the least bound neutrons, forming a skin-like surface layer, located predominantly at the nuclear periphery, extending to radii up to $r\approx 10$ fm. This can be clearly seen also from the examination of the wave function of the QRPA $2^+$ states in both $^{112,114}$Sn nuclei. All of the quadrupole QRPA states at E$_x$ = $2-5$ MeV have a predominant neutron structure of more than 98$\%$ which incorporates weakly bound two-quasiparticle neutron states (for more details see \,\cite{Tso11}). For comparison, we show also the neutron and proton QRPA transition densities of the well known collective isoscalar 2$^+_1$ state in these nuclei. 
Even though that the neutron excitations in the PQR region are dominating, the influence of the protons starts to increase with the decrease of the N/Z ratios. The effect is closely related with the decrease in the isovector interaction and the changes of the relative neutron and proton binding energies\,\cite{Tso11}. The latter influences the total PQR strength which increases with the increase of the neutron excess. 

The addition of quasiparticle-phonon coupling, as it is done in the EDF+three-phonon QPM approach \,\cite{Tso16}, induces more complex two-phonon and three-phonon configuration mixing into the 2$^+$ wave functions. The latter might be built of the same PQR-QRPA phonons but also from other phonons with higher energy and of different multipolarities, leading to additional isoscalar and isovector contributions to the electromagnetic transitions. The ($\alpha$,$\alpha'\gamma$) reactions act as a filter for the isoscalar components of the E2 strength but essentially leave undetected possible admixtures of isovector configurations. The EDF+QPM calculations, in fact, predict considerable contributions due to isovector components to the quadrupole transition matrix elements.
Consequently, the comparison between the experimental isoscalar B(E2) strengths obtained from the $\alpha$-scattering cross sections and the B(E2) strengths measured in ($\gamma$ ,$\gamma'$) scattering on $^{124}$Sn might indeed hint at an additional isovector admixture which is important to understand the complete experimental B(E2) strength. 

Two new experiments using inelastic proton-scattering with E$_p=$ 8 MeV followed by the coincident detection of $\gamma$-rays in $^{112}$Sn and $^{114}$Sn nuclei have been performed at the Institute for Nuclear Physics of the University of Cologne (Germany) to excite low-spin states in the two lightest stable Sn isotopes. In these studies, the combined spectroscopy setup SONIC@HORUS has been used to detect the scattered protons and the emitted $\gamma$-rays of the respective excited states in coincidence \cite{Pic17}. The novel ($p, p' \gamma$) DSA coincidence technique allowed for the measurement of several lifetimes, which were not known before \cite{Hen15,Spi18}.
In addition, $\gamma$-decay intensities to different final states were determined by applying excitation-energy gates. The latter also exclude feeding contributions when using the ($p,p'\gamma$) DSA coincidence technique.
 
From the measured $\gamma$-decay intensities the ground-state $\gamma$-decay branching ratio $b_0$ have been calculated by the relations: $b_0= \left(1+\sum_i\Gamma_i/\Gamma_0\right)^{-1}= \Gamma_0/\Gamma$, where $\sum_i\Gamma_i/\Gamma_0$ is the intensity of all observed $\gamma$-decays to excited states and normalized to the ground-state decay intensity and $\Gamma= \Gamma_0 + \sum_i\Gamma_i$ is the total observed $\gamma$-decay width, see also the recent work on $^{124}$Sn \cite{Spi16}. In addition branching ratios of $\gamma$-decays to the 2$^+_1$ state $b_1= \Gamma_1/\Gamma$ where $\Gamma_1$ is the $\gamma$-decay width to the first 2$^+$ state are derived from the present experiments. Firm spin-parity assignments were either known from previous experiments \cite{ENSDF} or are based on the observed $\gamma$-decay behavior seen in the present ($p,p'\gamma$) experiments. Tentative assignments are given for those cases where the observed $\gamma$-decays allow for both a dipole or quadrupole character of the observed transitions, see Figs. 2 and 3. Usually, 1$^-$ or 1$^+$ states at energies between 2-5 MeV have much shorter lifetimes than the 2$^+$ states seen here, i.e. $\tau$ $<$ 100 fs. So, we might conclude that based on the lifetimes for most of the tentative spin assignments in $^{112,114}$Sn, a 2$^+$ assignment is the more likely scenario. This conclusion is supported by the finding that in $^{124}$Sn, where spin assignments were possible, no 1$^-$ or 1$^+$ states were found with lifetimes comparable to 2$^+$ states. We, thus, exclude the excited state at 3873 keV in $^{112}$Sn and the excited states at 3933 keV and 4022.4 keV in $^{114}$Sn from the systematic shown here. For further details of the experiment, see Ref.\cite{Spi18}. 


Before we discuss our PQR results from the EDF+ three-phonon QPM theory and the experiment in $^{112,114}$Sn, we performed detailed studies of the lowest-energy 2$^+$ and 4$^+$ states and corresponding E2 transitions. In Table\,\ref{Tab1} the theoretical excitation energies and transition strengths are compared to experimental data of the current and previous experiments \,\cite{Spi18}, respectively. In general, the EDF+QPM calculation are able to describe the properties of low-lying 2$^+$ and 4$^+$ states. In addition, for the $^{114}$Sn nucleus the EDF+QPM results are compared to IBM-2 calculations from \,\cite{Spi18}. The observed deviations of $B(E2; 4^+_1\longrightarrow 2^+_1 )$ and $B(E2; 2^+_3 \rightarrow 2^+_1)$ transition values of the IBM-2 and the data could be explained in our approach the following way. In particular, the relative mixing between one- and two-phonon components in the structure of 2$^+_3$ and 4$^+_1$ states determines the transition rates to the lowest-lying 2$^+_1$ state in $^{112,114}$Sn nuclei. In both cases, the two-phonon $\left[2^+_1\otimes2^+_1\right]$ component of 2$^+_3$ and 4$^+_1$ states is of major importance for the value of the E2 transition matrix elements as far as it is related to the exchange of a collective quadrupole phonon. The EDF+QPM  predicts for $^{112}$Sn(2$^+_3$) a $4\%$ admixture of the two-phonon $\left[2^+_1\otimes2^+_1\right]$ component. In comparison, in $^{114}$Sn(2$^+_3$), with a slightly larger neutron excess, the $\left[2^+_1\otimes2^+_1\right]$ configuration contribution increases to $12\%$. Consequently, the $B(E2; 2^+_3 \rightarrow 2^+_1)$ transition probability is larger - 3 W.u. in $^{114}$Sn than in $^{112}$Sn where it is $\approx$1 W.u. which is also $\approx$3 times less as it is indicated by the structure of the states. Another interesting case is the 4$^+_1$ state in these nuclei. The experimental excitation energy of this state in $^{112}$Sn is larger than in $^{114}$Sn and it is nicely reproduced by the EDF+QPM theory. According to our theoretical findings the structure of the 4$^+_1$ states in both nuclei is predominantly, more than 70$\%$ of one-phonon character. Nevertheless, the contribution of the $\left[2^+_1\otimes2^+_1\right]$ component in these states plays an important role in the corresponding E2 transitions.  
The theoretical $B(E2; 4^+_1\longrightarrow 2^+_1 )$ transition probability in $^{112,114}$Sn is shown in Table\,\ref{Tab1}. Variations of the $\left[2^+_1\otimes2^+_1\right]$ component admixtures, on the level of few percent modify significantly the $B(E2; 4^+_1 \rightarrow 2^+_1)$ values of these nuclei. In this sense the investigated $B(E2; 2^+_3 \rightarrow 2^+_1)$ and $B(E2; 4^+_1\longrightarrow 2^+_1 )$ transitions are a clear evidence of the admixture of multi-phonon components in the nuclear excited states. In general, the EDF+QPM results are found in good agreement with the experiment as seen in Table\,\ref{Tab1}.

In order to examine in detail the low-energy quadrupole strength associated with the PQR in $^{112,114}$Sn, we compare the theoretical EDF+QPM and experimental B(E2) strengths, ground-state and 2$^+_1$ state $\gamma$-decay branching ratios, $b_0$ and $b_1$, respectively. We limited the considered final states in the EDF+QPM calculation of $b_0$ and $b_1$ to positive-parity final states which were also observed in the experiment. Those include states up to the 2$^+_3$ state as well as the 4$^+_1$ state. 

\begin{figure}[tb]
\centering
\includegraphics*[width=84mm]{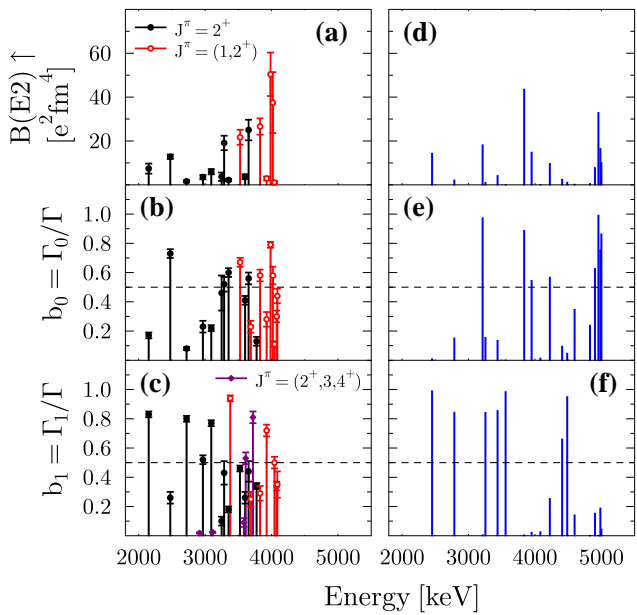}  
\caption{(color online)$^{112}$Sn: (a) Experimental E2 transition strengths, (b) ground-state $\gamma$-decay branching ratio $b_0$ and (c) 2$_1^+$ state $\gamma$-decay branching ratio $b_1$ in the energy range $E_x=2 - 4.2$ MeV; (d) EDF+QPA E2 strengths, (e) ground-state $\gamma$-decay branching ratio $b_0$ and (f) 2$_1^+$ state $\gamma$-decay branching ratio $b_1$ in the energy range $E_x=2 - 5$ MeV. The horizontal dashed line shows b$_0$ = 0.5 and b$_1$=0.5. See text for further details.}
\label{FIG2}
\end{figure}

\begin{figure}[tb]
\centering
\includegraphics*[width=84mm]{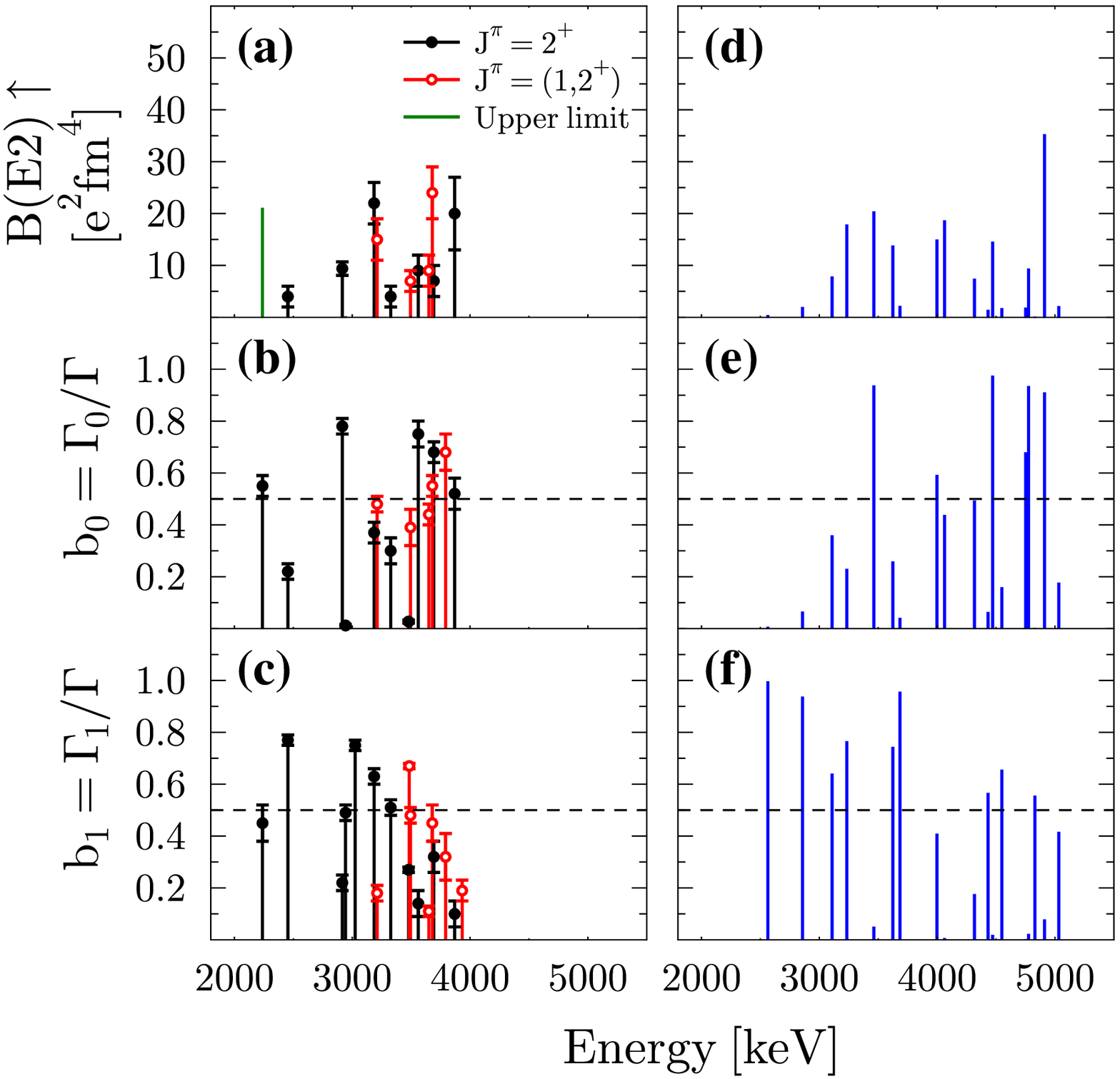}  
\caption{(color online)$^{114}$Sn: (a) Experimental E2 transition strengths, (b) ground-state $\gamma$-decay branching ratio $b_0$ and (c) 2$_1^+$ state $\gamma$-decay branching ratio $b_1$ in the energy range $E_x=2 - 4.2$ MeV; (d) EDF+QPA E2 strengths, (e) ground-state $\gamma$-decay branching ratio $b_0$ and (f) 2$_1^+$ state $\gamma$-decay branching ratio $b_1$ in the energy range $E_x=2 - 5$ MeV. The horizontal dashed line shows b$_0$ = 0.5 and b$_1$=0.5. See text for further details.
}
\label{FIG3}
\end{figure}

\begin{figure}[tb]
\centering
\includegraphics*[width=84mm]{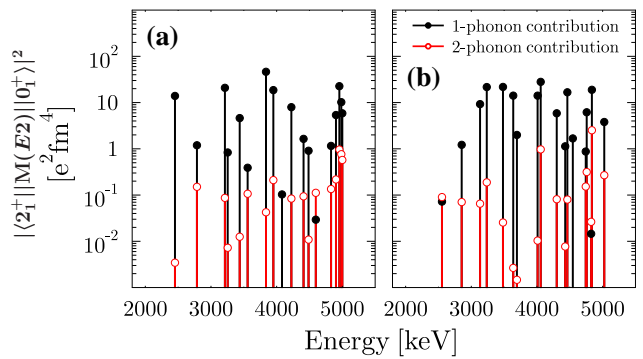}  
\caption{(color online) Calculated EDF+QPM summed contributions of one- and two-phonon squared E2 matrix elements of ground state decay transitions of PQR states in (a) $^{112}$Sn and (b) $^{114}$Sn.
}
\label{FIG4}
\end{figure}

Experimental and EDF+QPM spectral distributions of 2$^+$  states and corresponding branching ratios, $b_0$ and $b_1$ of $^{112}$Sn and $^{114}$Sn are shown in Fig.\,\ref{FIG2}(a), (b), (c), (d), (e) and (f) and Fig.\,\ref{FIG3}(a), (b), (c), (d), (e) and (f), respectively. 
As in our previous PQR investigations in $^{124}$Sn\,\cite{Spi16}, we observe a sequence of low-energy 2$^{+}$ states which decay predominantly to the ground state. Such states are characterized by large g.s. branching ratios, typically $b_0\geq 50\%$ ; see Fig.\,\ref{FIG2}(b,e) and Fig.\,\ref{FIG3}(b,e). 
In $^{112,114}$Sn this sequence of quadrupole excitations involves nine almost pure neutron QRPA states of different collectivity which could be associated with a PQR mode and neutron skin oscillations. Furthermore, the PQR one-phonon states are a substantial part of all QPM 2$^+$ states at E$_x$= $2-5$ MeV. In this respect, the theoretical analysis of the QPM $2^+$ excitations at $E_x\approx 2-5$ MeV in $^{112,114}$Sn isotopes confirms the expectations of a PQR mode in nuclei with relatively small neutron excess. In particular, the 2$^+$ states at $E_x\approx 2-4$ MeV have a predominant one-phonon structure, more than $50\%$ given mainly by the PQR-QRPA 2$^+$ states at $E_x\approx 2-5$ MeV. An exception is the QPM 2$^+_2$ state in $^{112}$Sn which contains a major two-phonon $\left[2^+_1\otimes2^+_1\right]$ component. At excitation energies E$_x$= $4-5$ MeV the increase of the level densities leads to stronger fragmentation of the single-particle PQR strength. As a results, in both tin nuclei the QPM 2$^+$ states at E$_x$= $4-5$ contain large two-phonon counterparts. Nevertheless, how small it may be, the one-phonon admixture in these 2$^+$ states is most important for their g.s. transitions. Consequently, the one-phonon PQR configurations give much larger  contributions to the E2 transition matrix elements $\left\langle 2^+_i\left\|M(E2)\right\|g.s.\right\rangle$ for most of the QPM 2$^+$ states at $E_x\approx 2-5$ MeV than the two-phonon components. An exception are two 2$^+$ QPM states where the one- and two-phonon contribution to the M(E2) value are of almost the same size. In Fig.\,\ref{FIG4}, the  one- and two-phonon contributions to the PQR transition strengths are shown.

A strong evidence supporting the admixture of one-phonon and multi-phonon configurations in the structure of PQR states could be obtained from investigations of $b_0$ and $b_1$ branching ratios.
Commonly, the QPM PQR states have $b_0$ values close or well above 50$\%$ and they prefer decaying to the ground state than to an excited state. However, there are also exceptions which mainly concern the presence of the two-phonon $\left[2^+_1\otimes2^+_1\right]$ component in the structure of the 2$^+$ states. Considerable contributions of three-phonon configurations to the structure of 2$^+$states are found at energies above 4 MeV in both, $^{112}$Sn and $^{114}$Sn nuclei.

Thus, in 2$^+$ states with two-phonon admixtures which contain the collective $[2_1^+]_{QRPA}$ phonon, the strong E2 transitions to the 2$^+_1$ lead to the reduction of the $b_0$ value.
As far as the observation of large $b_0$ values is a signature of the major one-phonon contribution to excited 2$^+$ states, the observation of large $b_1$ values is a clear indication for two-phonon component in these states. At the same time, excited states with larger $b_0$ values have smaller $b_1$ values and vice versa. This is clearly seen in Fig.\,\ref{FIG2}(b), (c), (e) and (f) and Fig.\,\ref{FIG3}(b), (c), (e) and (f), respectively. 

The results emphasize that the investigations of branching ratios are playing an important role for nuclear research.
In particular, branching ratios serve as a highly sensitive observable for  distinguishing simple particle-hole type configurations and multi-phonon structures.  
The properties of the PQR mode are revealed by investigating the fragmentation pattern of the PQR QRPA $2^+$ states over the excited $2^+$ states, both experimentally and theoretically by the EDF+three-phonon QPM calculations.

In fact, the general trend predicted by the EDF+QPM theory on B(E2) transition strengths and corresponding branching ratios is in good agreement with the experimental observations for both tin isotopes. The calculated result within EDF+QPM for the total PQR strength, $E_x\approx 2-5$ MeV is B(E2)$_{QPM}^{PQR}$= 179 e$^2$fm$^4$ in $^{112}$Sn and B(E2)$_{QPM}^{PQR}$= 188 e$^2$fm$^4$ in $^{114}$Sn, respectively. For comparison the corresponding pure one-phonon PQR strengths obtained from our EDF+QRPA calculations is B(E2)$_{QRPA}^{PQR}$= 111 e$^2$fm$^4$ in $^{112}$Sn and B(E2)$_{QRPA}^{PQR}$= 120 e$^2$fm$^4$ in $^{114}$Sn, respectively. The experimental summed B(E2) strength between 2 MeV and 4.2 MeV is 224(19) e$^2$fm$^4$ in $^{112}$Sn and 151(16) e$^2$fm$^4$ in $^{114}$Sn. For firm assignments these strengths are 84(7) e$^2$fm$^4$ and 97(14) e$^2$fm$^4$, respectively. As mentioned above, three states have been excluded based on their lifetime being atypical for 2$^+$ states at these energies. The B(E2;$0^+_1 \rightarrow 2^+_i$) reduced transition probability for the states at 3933 keV and 4022 keV in $^{114}$Sn would be 184(88) e$^2$fm$^4$ and 215(60) e$^2$fm$^4$, respectively.

In conclusion, EDF plus three-phonon QPM calculations of the low-energy electromagnetic quadrupole response of $^{112,114}$Sn nuclei are compared to two new high resolution ($p$, $p' \gamma$) Doppler-shift attenuation (DSA) coincidence experiments and  permit an identification of a new mode of nuclear excitation in those two nuclei, i.e. the pygmy quadrupole resonance (PQR). The origin of the PQR is explained suggesting two-quasiparticle neutron excitations related to neutron skin oscillations. The EDF calculations of neutron skin thickness in $^{112,114}$Sn indicate an increase of the calculated total PQR strength with the neutron excess.
This is further confirmed in the studies of neutron and proton transition densities which indicate an isoscalar and isovector contributions to PQR excited states. Furthermore, a substantial contribution from the low-energy multi-phonon states induces considerable additional E2 strength in the PQR region. From these studies it is also observed that the low-energy PQR strength is strongly fragmented. 
A detailed information of the fine structure of PQR mode could be obtained from branching ratios of ground and excited states which are observed, both theoretically and experimentally.
In this aspect, both b$_0$ and b$_1 $ branching ratios, serve as a sensitive indicator of  the microscopic structure of nuclear excitations. Furthermore, the theoretical observations show that branching ratios give useful information on the collectivity of the excited states. In particular, non-collective 2$^+$ states typically have smaller b$_0$ branching ratios. However, due to the  admixture of collective two-phonon configurations such states might have larger b$_1$ branching ratios. 

Thus, the theoretical and experimental results of E2 transitions and branching ratios in $^{112,114}$Sn contribute to the understanding of the fine structure of the PQR and give a strong evidence on the existence of multi-phonon excitations in the PQR energy region. In general, the EDF+QPM theory successfully reproduces the E2 spectral distributions, both b$_0$ and b$_1 $ branching ratios and provides a very good description of the total B(E2) transition probabilities of the low-energy 2$^+$ and 4$^+$  states observed in the experiment. 
The work reported here illustrates the vital role of combined theoretical and experimental efforts in revealing the complex nature of exotic modes and the fine structure of nuclear excitation in neutron-excess nuclei.

This work was performed under the auspices of DFG grant ZI 510/7-1. The authors would like to thank P. Petkov for his support with the Monte-Carlo simulation code used for the lifetime determination. Furthermore, we gratefully acknowledge the help of the accelerator staff at the IKP Cologne.

\vfill\eject

\end{document}